\documentclass[11pt]{article}

\usepackage{float}

\usepackage{comment}

\usepackage[tmargin=2cm, lmargin=2.5cm, bmargin=3cm,hmarginratio=1:1,headheight=65pt,headsep=1cm]{geometry} %hmarginratio est par dÃ©faut 2:3 (pages impaires/paires)
\usepackage[T1]{fontenc}
\usepackage[french,english]{babel}
\usepackage[latin1]{inputenc}%Toma acentos%

\usepackage{color,graphicx}
\usepackage{amsmath,amssymb,amsfonts}
\usepackage{amsthm}
\usepackage{mathtools}
\usepackage{slashed}
\usepackage{sidecap} % caption to the right of the figure
\usepackage{array}
\usepackage[indentafter]{titlesec}%indentation en dÃ©but de section

\usepackage{setspace}
\usepackage{perpage} %the footnote perpage package
\usepackage{fancyhdr} % for the style of the headers
\usepackage{emptypage}% so that no headers appear on blanck pages
\usepackage{pifont} % for more symbols

\usepackage{hyperref} % SHALL ALWAYS BE THE LAST PACKAGE
\usepackage{color}

\newcommand{\pink}[1]{\textcolor{\pink}{#1}}

\definecolor{dblue}{rgb}{0.2,0.50,0.80}

\newcommand{\g}{\g}
\newcommand{\p}{\partial}

\newcommand{\be}{\begin{eqnarray}}
\newcommand{\en}{\end{eqnarray}}
\newcommand{\badat}{\begin{alignedat}}
\newcommand{\eadat}{\end{alignedat}}
\newcommand{\bitm}{\begin{itemize}}
\newcommand{\eitm}{\end{itemize}}
\newcommand{\bmat}{\begin{pmatrix}}
\newcommand{\emat}{\end{pmatrix}}
\newcommand{\ba}{\begin{align}}
\newcommand{\bas}{\begin{align*}}
\newcommand{\ab}{\end{align}}
\newcommand{\bse}{\begin{subequations}}
\newcommand{\ese}{\end{subequations}}

\newcommand{\ee}{\hspace{5 mm}}

\def\D{\mathcal{D}}

\def\L{\mathcal{L}}

\def\S{\mathcal{S}}
\def\g{ \gamma}

\def\b{\theta(\phi)}
\def\cN{{\cal N}}

\def\be{\begin{equation}}
\def\ee{\end{equation}}
\def\bea{\begin{eqnarray}}
\def\eea{\end{eqnarray}}
\def\ba{\begin{array}}
\def\ea{\end{array}}
\def\bec{\begin{center}}
\def\ec{\end{center}}
\def\ba{\begin{align}}
\def\ena{\end{align}}

\def\lab{\label}

\def\12{\frac{1}{2}}

\def\a{\alpha}
\def\b{\beta}
\def\g{\gamma}

\def\d{\delta}
\def\D{\Delta}

\def\i{\iota}

\def\l{\lambda}
\def\L{\Lambda}
\def\m{\mu}
\def\n{\nu}

\def\r{\rho}

\def\s{\sigma}
\def\t{\tau}

\def\lp{\left(}
\def\rp{\right)}
\def\lbr{\left[}
\def\rbr{\right]}

\def\S{\Sigma}

\begin{document}

\title{\textbf{Thin-shell wormholes in AdS$_5$ and string dioptrics}}

\author{Mariano Chernicoff$^1$, Edel Garc\'ia$^1$, Gaston Giribet$^2$, Emilio Rub\'{\i}n de Celis$^2$}

\date{}
\maketitle

\begin{center}

$^1${ Departamento de F\'{\i}sica, Facultad de Ciencias, Universidad Nacional Aut\'{o}noma de M\'{e}xico}\\
{{\it A.P. 70-542, CDMX 04510, M\'{e}xico.}}

\smallskip

$^2${ Physics Department, University of Buenos Aires and IFIBA-CONICET}\\
{{\it Ciudad Universitaria, pabell\'on 1 (1428) Buenos Aires, Argentina.}}

\smallskip

\smallskip

\smallskip

\end{center}

%%%%%%%%%%%%%%%%%%%%%%%%%%%%%%%%%%%%%%%%%%%%%%%%%%%%%%%

\begin{abstract}
We consider string probes in a traversable wormhole geometry that connects two locally AdS$_5$ asymptotic regions. Holographically, this describes two interacting copies of a 4-dimensional gauge theory. We consider string configurations whose endpoints are located either in the same boundary or in the two different boundaries of the wormhole. A string with both endpoints in the same boundary is dual to a quark-antiquark pair charged under the same gauge field, while a string extending through the wormhole describes a pair of colored particles charged under two different gauge fields. When one considers a quark-antiquark pair in each boundary, the system undergoes a phase transition: While for small separation each pair of charges exhibits Coulomb interaction, for large separation the charges in different field theories pair up. This behavior had previously been observed in other geometric realizations such as locally AdS$_5$ wormhole solutions with hyperbolic throats. The geometries we consider here, in contrast, are stable thin-shell wormholes with flat codimension-one hypersurfaces at fixed radial coordinate. They appear as electrovacuum solutions of higher-curvature gravity theories coupled to Abelian gauge fields. The presence of the thin-shells produces a refraction of the string configurations in the bulk, leading to the presence of cusps in the phase space diagram. We discuss these and other features of the phase diagram, including the analogies and difference with other wormhole solutions considered in related contexts.
\end{abstract}

\maketitle

\vspace{-1cm}
%\tableofcontents

\newpage

\section{Introduction}

AdS/CFT correspondence \cite{Malda, GKP, Witten} provides a powerful framework to study the dynamics of strongly coupled systems and, in particular, gauge theories. A very interesting scenario where AdS/CFT has proven to be very useful is the so-called quark-gluon plasma (QGP). This state of matter, for a brief period of time does not behave as a weakly coupled gas of quarks and gluons, but rather as a strongly coupled fluid \cite{fluid,fluid2}. Partly motivated by the idea of making contact with experimental data obtained from RHIC and LHC \cite{alice, qgprev}, and assuming that a real-world QCD can be reasonably well approximated by at least one of the various `QCD-like' gauge theories whose dual description is known, a few years ago a new interesting field of research begun: Starting with the seminal work \cite{Policastro}, where the shear viscosity of the QGP was computed, it begun a long list of different fruitful applications and a variety of techniques that allowed to explore different aspects of such an interesting system. One such line of research, for example, focused on calculating the drag force experienced by a heavy quark traversing an $\mathcal{N}=4$ super-Yang-Mills plasma via its dual description as a string moving on an AdS-Schwarzschild background  \cite{Herzog, Gubser, CT}. Closely related, the dynamics of mesons moving through this plasma were studied in \cite{Liu, Mariano1}, where the corresponding quark-antiquark potential and screening length were obtained using the dual portrayal in terms of a string that has both of its endpoints on the boundary. The standard description states that the string endpoints represent the quark and antiquark, while the body of the string represents the color `flux tube' in between, i.e. the gluonic field profile sourced by the fundamental color sources. The particular case where the quark and antiquark were static in the thermal %\footnote{In a different but related context, it is also worth mentioning a series of papers where thermalization was studied from the holographic point of view \cite{T1, T2, T3, T4, T5, T6, T7, T8, T9, T10, T11, T12}.}
medium was previously analyzed in \cite{theisen, brandhuber}. Another important line of research focused on studying the behavior of these probes when the dual conformal field theory is at zero temperature. Such scenario is interesting on its on right because, being a strongly coupled gauge theory, and assuming that some properties might be universal among seemingly different gauge theories, learning from these complicated systems, though different from QCD, could be very useful. For example, and closely related to what we will study in this paper, the static potential for an infinitely heavy quark and antiquark pair was first computed in \cite{Rey}, where, as expected, due to conformal invariance, for small separation of the pair, the potential was found to be proportional to $1/L$ (where $L$ represents such separation). In a similar context, another important problem that has been addressed is the dynamics of a heavy color-charged particle in vacuum. The first characterization of the radiation rate of an accelerating quark was worked out in \cite{mikhailov}, where, remarkably, the author was able to solve the highly nonlinear equation of motion for a string on an AdS$_5$ background that follows a timelike trajectory of the string endpoint dual to an infinitely massive quark. In \cite{lorentzdirac, lorentzdiraclargo, unruh}, a careful analysis of different dynamical physical situations were studied giving a nice and clear picture of energy loss in strongly coupled CFT's. 

Partly motivated by recent developments in the AdS/CFT correspondence and its ramifications, more precisely by the picture presented in \cite{Maldacenasusskind}, one could be interested in probing wormhole geometries connecting two asymptotic AdS regions using fundamental strings as those described above. Some work along these lines has been done in \cite{Vazquez1, Vazquez2, Takayanagi, Raulo, Vazquez3, Vazquez4}, where different string configurations were considered on a two-sided AdS$_5$ traversable wormhole. In that case, the holographic dual is given by two copies of the gauge theory which are interacting. In \cite{Vazquez1}, for example, the authors studied the dynamics of open strings on a geometry with two asymptotic regions, each of them being locally equivalent to AdS$_5$ or, more precisely, to the universal covering of AdS$_5$ with constant-radius slices of negative constant curvature. This type of traversable wormhole geometry, with a base manifold of the form $\Sigma_3 = \mathbb{H}_3/\Gamma_3$ or $\Sigma_3 =S^1\times \mathbb{H}_2/\Gamma_2 $, with $\Gamma _{2,3}$ being a discrete subgroup of $\mathbb{H} _{2,3}$, appears as solutions of higher-curvature theories such as Einstein-Gauss-Bonnet (EGB) \cite{Julio}. A string solution with both endpoints on the same side of the wormhole describes a pair of colored charges within the same copy of the gauge theory, and exhibit Coulomb interaction for small separation. A string extending through the wormhole, in contrast, describes a pair of charges which live in different copies of the gauge theory, and they exhibit a spring-like potential. As shown in \cite{Vazquez1, Vazquez2}, a phase transition occurs when there is a pair of charges present within each field theory. In that case, for small separation each pair of charges exhibits Coulomb interaction; however, for large separation the charges in the different field theories pair up. 

The purpose of this work is to analyze similar string configurations on two-sided wormhole geometries, but in a setup that is more natural from the holographic point of view; namely, considering geometries with flat constant-radius slices instead of hyperbolic higher-genus ones. The reason being that in the former case the gauge theory is formulated on a locally flat 4-dimensional spacetime. We want to see whether the same features, such as the existence of phase transitions, are also possible in the flat sliced AdS$_5$ wormhole case. There is, however, an obvious obstruction to achieve this: No-go theorems seem to forbid the existence of such wormhole configurations precisely in those gravity theories in which, as in EGB, the problem is under control analytically. Even at the so-called Chern-Simons point of the theory \cite{Zanelli}, where the Birkhoff-like no-go theorems can be evaded \cite{Zeger}, the space of solutions only contains differentiable static wormholes with hyperbolic slices \cite{Julio}. This compels us to get rid of differentiability condition and to consider thin-shell wormholes of the type studied in \cite{Garraffo}; that is, thin-shell electro-vacuum solutions of EGB theory \cite{elanterior}.

EGB theory, also known as 5D Lovelock theory \cite{Lovelock1, Lovelock2}, has shown to be the best suitable toy model to investigate the effects of higher-curvature effects, especially in the context of AdS/CFT \cite{Brigante1, Brigante2, Buchel, Hofman, Li, Zeng, Jose1, Jose2, Jose3, Jose4, Jose5, Jose6, Maldacena}. This was the model considered, for example, in \cite{Brigante1, Brigante2} to investigate the shear viscosity to entropy density ratio in conformal field theories dual to Einstein gravity with curvature square corrections. Yielding second-order field equations, EGB theory enables to compute the shear viscosity non-perturbatively in the dimensionful coupling constant of the higher-curvature terms. The possibility of performing such computation analytically is what ultimately made possible for the authors of \cite{Brigante1, Brigante2} to show that the previously conjectured Kovtun-Son-Starinet viscosity bound \cite{KSS} was violated when higher-curvature terms were present.

Here, we will consider EGB theory and the wormhole solutions constructed in \cite{elanterior}. These geometries are electro-vacuum solutions to EGB gravity coupled to an Abelian gauge field, with an electric flux that suffices to support the wormhole throat and to stabilize the geometry. In \cite{elanterior} we presented a protection argument showing that, while stable traversable wormholes connecting two asymptotically locally AdS$_5$ regions do exist in EGB theory, the region of the parameter space where such solutions are admitted lies outside the causality bounds coming from AdS/CFT \cite{Brigante1}; see (\ref{Unitaritybound}) below. There exists, however, a special case: For a special relation between the coupling constant of the higher-curvature terms ($\alpha $) and the cosmological constant ($\Lambda $), which are defined in (\ref{Uno}) below, the argument used to exclude a point of the unitarity segment simply does not hold. To be more precise, if the coupling constants satisfy the relation
\begin{equation}
\alpha \Lambda \, = \, -\frac{3}{4}\ , \label{CSpoint}
\end{equation} 
then the EGB gravity theory in vacuum reduces to the 5D Chern-Simons (CS) theory for the gauge group $SO(4,2)$ \cite{Zanelli} and, in that case, no graviton excitation exist around AdS$_5$. As a consequence, the argument leading to the unitarity bound
\begin{equation}
 -\frac{27}{100} \, \leq \, \alpha \Lambda \, \leq \, \frac{7}{12}\ , \label{Unitaritybound}
\end{equation} 
does not suffice to exclude the theory at (\ref{CSpoint}), which stands as an isolated point outside the segment: the Chern-Simons point. It is worth mentioning that this is exactly the same point where the wormhole solutions considered in \cite{Vazquez1, Vazquez2} exist. Here, we will also study the theory at this point (\ref{CSpoint}), however, we will focused on geometries such as those constructed in \cite{elanterior}. The paper is organized as follows: In section 2, we review the construction of the traversable thin-shell wormholes in 5D CS gravity. The higher-curvature terms will suffice to satisfy the junction conditions on the thin-shells with no induced energy-momentum tensor on the constant-radius codimension-1 separation hypersurfaces. In section 3, we will solve the string equations of motion on such background. Interestingly, as the the string probes the deep IR of the wormhole geometry, the latter, being non-differentiable on the thin-shell, will manifest itself as a `lense', changing the shape of the string worldsheet and producing a refraction effect: a `string refraction'. The dynamics of the string dioptrics is studied in section 4, where, in particular, the phase transition is discussed along with the different possible string configurations. In section 5, we compare our string configurations with those previously studied on traversable wormholes in a two-sided AdS$_5$ space.

\section{Traversable wormholes}
\lab{intro}

\subsection{The higher-curvature theory}

Higher-curvature gravity models in 5 dimensions do allow for stable wormholes that connect two asymptotically flat or asymptotically (Anti-)de Sitter regions without introducing extra exotic matter, \cite{elanterior}: We consider the Einstein-Maxwell theory supplemented with the quadratic Gauss-Bonnet higher-curvature terms; namely
\begin{equation}
S_5 =\frac{1}{16\pi }\int \, d^5x\, \sqrt{-G}\, \Big(\, R-2\Lambda +{\alpha}\, \Big( R_{\mu \nu \rho \sigma }R^{\mu \nu \rho \sigma }-4R_{\mu \nu }R^{\mu \nu }+R^2\Big)
-\,  F_{\mu \nu} F^{\mu \nu}\, \Big) +B \ ,
\label{Uno}
\end{equation}
where $B$ stands for the boundary term that renders the variational problem well defined.  

The static black hole solution of this theory can be written analytically in a closed form \cite{Boulware:1985wk, Wiltshire:1988uq}, namely
\be \lab{metric}
ds^2 = \, G_{\mu \nu }\, dx^{\mu }dx^{\nu } = - N^2(r)\, f(r)\, dt^2 + \frac{dr^2}{f(r)} + r^2 d\Sigma_3^2 \ ,
\ee
where $t\in \mathbb{R}$, $r\in \mathbb{R}_{\geq 0}$, and $d\Sigma_3^2$ is the metric on a locally, maximally symmetric 3-space of constant curvature $k=0,\pm 1$. The metric functions are given by
\be
N^2=1\ , \ \ \, f(r) = k + \frac{r^2}{4 \a}  \lp 1 + \xi  \sqrt{1 +  \frac{16 \a M}{r^4} -  \frac{8 \a Q^2}{3 r^6} + \frac{4 \a \Lambda}{3}} \rp \lab{BD} \ ,
\ee
where $\xi =\pm 1$. The electromagnetic field, on the other hand, is given by 
\begin{equation}
F  \, = \, F_{\mu \nu }\, dx^{\mu }\wedge dx^{\nu }  \, =  \, \frac{Q}{r^3}\, \, dt\wedge dr\, ,\label{KKKK}
\end{equation} 
with $Q$ being the electric charge.

For certain range of parameters, solution (\ref{metric})-(\ref{KKKK}) represents a static, electrically charged black hole with ADM mass proportional to $M$. Here, we will be interested in solutions with flat base manifold, namely $k=0$, and in the theory defined at the point (\ref{CSpoint}). In that case, (\ref{BD}) reduces to 
\be \lab{BD2}
f(r) = \frac{r^2}{4 \a}   +   {\xi}\sqrt{ \frac{ M}{\alpha } -  \frac{ Q^2}{6\alpha r^2} } \, , \ \ \ \  \ d\Sigma_3^2 =  d \vec{x}\cdot d\vec{x} = \delta_{ij}dx^idx^j  \, ,  
\ee
with $i,j=1,2,3$. The radius of the black hole event horizon is given by the largest positive root of the equation $r_H^6 - 16\alpha Mr_H^2+8\alpha Q^2/3=0$, for the solution with $\xi=-1$. When the radicand in (\ref{BD2}) vanishes, the solution exhibits a branch singularity. This singularity, which occurs at $r_S^2=Q^2/(6M)$, may be shielded by the event horizon provided $|Q|$ is smaller than a critical value $Q_c$ that guarantees $r_H >r_S$. If there are no horizons, a naked singularity is present at the radius where the radicand in (\ref{BD2}) vanishes, i.e. at the hypersurface $r=r_S$.

\subsection{Thin-shell wormholes at the Chern-Simons point}

We consider $\a>0$ and $\Lambda <0$ obeying (\ref{CSpoint}). In this theory, we will use solutions that, at large $r$, asymptote the spherically symmetric geometry described above with $\xi=-1$ and $k=0$, the latter implying the constant-$r$ slices being locally $\mathbb{R}^{1,3}$. More precisely, we will consider traversable wormhole solutions of the type studied in \cite{elanterior}, connecting two asymptotically locally AdS$_5$ regions. The geometric surgery procedure to construct the wormhole involves three different junctures, gluing four different patches of the geometry (\ref{metric}) with $\L =  -3/(4\a)$ and different values of $\xi $ and $M$ in each patch. The three different junctures describe vacuum thin-shells: The contribution of the higher-curvature terms, which has its counterpart in the boundary action, makes it possible to satisfy the junction conditions without induced matter on the shells. 

Let us be more precise about the geometric construction: The spacetime consists of four bulk pieces, four distinct patches separated by three 4-surfaces. Each piece will be equipped with a metric of the form (\ref{metric})-(\ref{BD2}), cutting out part of the geometry on one side of some codimension-one hypersurface defined at fixed coordinate $r$. The four bulk regions are then glued in such a way that the entire space $\mathcal{M}$ results geodesically complete and describes a traversable wormhole, symmetric with respect to one of the three thin-shells, the one that corresponds to the wormhole throat. We denote the four regions as left-exterior $\mathcal{M}_L^{e} $, right-exterior $\mathcal{M}_R^{e}$, left-interior $\mathcal{M}_L^{i}$, and right-interior $\mathcal{M}_R^{i} $, with coordinates ${x^{\mu}_{L,R}}$, where the subscripts $L$ and $R$ stand for Left and Right respectively. The wormhole is symmetric across the throat that separates the region $\mathcal{M}_L^{i}$ from the region $\mathcal{M}_R^{i}$; nevertheless, the coordinates of the interior and exterior patches could be different a priori. Without loss of generality, we will consider the same coordinates for the interior regions, and other coordinates system for the exterior ones. Explicitly, these regions are
\be
\begin{array}{lll} 
\mathcal{M}_L^{e} = \{ x^\mu_L \, / \, r_L \ge b \}, 
&\qquad&
\mathcal{M}_R^{e} =\{x^\mu_R \, / \, b \le r_R \}\,, \\ \\
\mathcal{M}_L^{i} = \{ x^\mu_L \, / \,  b \ge r_L \ge a \},
&\qquad&
\mathcal{M}_R^{i}=\{x^\mu_R \, / \,  a \le r_R \le b \},
\end{array}
\ee
and the complete manifold gets defined by $\mathcal{M}=\mathcal{M}_L^{e}  \cup \mathcal{M}_L^{i}  \cup \mathcal{M}_R^{i}  \cup \mathcal{M}_R^{e} $. Each external region, $\mathcal{M}_{L,R}^e$, is joined with its corresponding inner region, $\mathcal{M}_{L,R}^i$, at the hypersurface of a vacuum thin-shell (bubble) $\Sigma_{L,R}^b$. That is, $\Sigma_{L,R}^b=\mathcal{M}_{L,R}^i \cap \mathcal{M}_{L,R}^e$. The Left bubble is placed at $r_L=b$, while the Right bubble is placed at $r_R=b$. Inner regions are also glued to each other at the throat, located at $r_L = a = r_R$, which implies that $a < b$; namely $\Sigma^a=\mathcal{M}_{L}^i \cap \mathcal{M}_{R}^i$. Summarizing, the junction hypersurfaces are described as $\Sigma_L^b \equiv \partial \mathcal{M}_L^{e}|_{r=b}  = \partial \mathcal{M}_L^{i}|_{r=b}$ and $\Sigma_R^b  \equiv \partial \mathcal{M}_R^{e}|_{r=b}  = \partial \mathcal{M}_R^{i}|_{r=b}$, which define bubbles, and $\Sigma^a \equiv \partial \mathcal{M}_L^{i}|_{r=a}  = \partial \mathcal{M}_R^{i}|_{r=a}$, corresponding to the wormhole throat. 

%----

The metric functions $f_i(r_{L,R})$ and $f_e(r_{L,R})$ of the interior and exterior geometries, have mass parameters $M_{i} \equiv M_{Li} = M_{Ri}$ and $M \equiv  M_{Le} = M_{Re}$, respectively. As described above, the exterior metrics have $\xi_{e} = -1$ and asymptote a locally AdS$_5$ space at large $r$, with an effective AdS radius given by $\ell \equiv \sqrt{4\alpha}$. As analyzed previously in \cite{Garraffo,elanterior}, the juncture conditions on the hypersurface at $r=a$ determine that the inner regions must both have $\xi_{i} = +1$ branch to construct the symmetric wormhole. The radius of the throat has to be greater than the radius where the branch singularity exist, namely $a>r_{S_i}$. Also, the radii $r_{L,R} = b$ of the bubbles have to be greater than the would be horizons of the external metrics or their singularity surfaces, namely $b > \mbox{Max}\{ r_{S_e}, \,r_H \} $. The electric flux traversing the wormhole, given by the charge $Q \equiv -Q_L = Q_R$, suffices to support the wormhole throat and to stabilize the solution under scalar perturbations; see \cite{elanterior} for details; see also figure 1.

The general equations for the junctures at the shells of our construction are summarized in equations (21)-(24) % or (14)-(17)
of \cite{elanterior}. These are four equations which determine the conditions to have vacuum shells, i.e. a vanishing induced stress-tensor at the hypersurfaces.
The conditions are radically simplified for flat base manifold, $k=0$, and by setting the CS point $\L=-3/\ell^2$. 
After applying the junction conditions, four parameters of the metric functions are fixed, while other two remain free. We can present the thin-shell wormhole explicitly, described in terms of the radius of the throat $a$ and of the AdS radius $\ell$; namely 
\be
ds^2 = - f \, dt^2 + f^{-1} \, dr^2 + r^2 \, d\vec{x}^{\,2} \ ,
\ee
where
\be
f =
   \begin{dcases}
     f_i(r) = \frac{r^2}{\ell^2} +  \frac{2\,a^2}{\ell^2}\,   \sqrt{3  - \frac{2\,a^2}{r^2}}  
     &\mbox{in $\mathcal{M}_{L,R}^{i}$}
     \\
     f_e(r) =
\frac{r^2}{\ell^2} -  \frac{2\,a^2}{\ell^2}\,   \sqrt{ \bar{\mu}  - \frac{2\,a^2}{r^2}} 
\quad (\bar{\mu} \simeq 1.81)\,
&\mbox{in $\mathcal{M}_{L,R}^{e}$} 
 \, 
   \end{dcases}
\ee
with the corresponding coordinates on each patch. The parameters of the original metric functions, obtained by the junction conditions, are
\begin{align} 
\label{Mi_Q}
M_i &=   \frac{3\, a^4}{\ell^2} \; 
&&Q =  \frac{2 \,\sqrt{3} \, a^{3}}{\ell} \; &
\\
\label{M_b}
M  &= \frac{ \bar{\mu} \, a^4}{\ell^2} \, \qquad (\bar{\mu} \simeq 1.81) \,,
 &&\; b = \sqrt{\d} \, a \, \qquad (\sqrt{\d} \simeq 1.40) \, &
\end{align}
with the exact value of $\d$ given by the positive root of the polynomial $P( \d ) =  9\d^7  - 21 \d^6  + 4 \d^5 + \d^4  + 16 \left( \d^3 - \d^2 - \d - 1 \right)$, and with $\bar{\mu} =  10/\d  - 3 - 4 /\d^{4}$. The mass parameter of the interior metric, and the electric charge of the wormhole, given in (\ref{Mi_Q}), are obtained from the equations for the junction at the throat (equations (21) and (22) in \cite{elanterior}). Plugging the latter parameters into the conditions on the bubbles (equations (23) and (24) in \cite{elanterior}) we have the radius of bubbles and the external mass parameter, as given in (\ref{M_b}).

The solutions have supracritical values of $Q$ for the outer geometries: The exterior function $f_e(r)$ has a charge $Q$ such that $Q_c/Q = \sqrt{2} \, (\bar{\m}/3)^{3/4} \simeq 0.97$. We also notice that the branch singularities of the metric functions are excluded from the wormhole geometry; the outer function $f_e(r)$ has $r_{S_e}/b = \sqrt{2}/\sqrt{\d \bar{\m}} \simeq 0.75$, while the inner function have $r_{S_i}/a = \sqrt{2/3} \simeq 0.81$, as given by the radicand in $f_i(r)$. We emphasize that the construction is stable under scalar perturbations. The latter can be seen from the stability equations for the throat and bubbles found in \cite{elanterior}. There are no horizons and the two asymptotic regions are causally connected: a radial null geodesic extends from one boundary to the other in a finite time. In this way we have a traversable asymptotically locally AdS$_5$ wormhole spacetime 
%with flat base manifold and 
with both boundaries being locally $\mathbb{R}^{1,3}$. 

%%%%%%%%%%%%%%%%%%%%%%%%%%%%%%%%
%%%%%%%%%%%%%%%%%%%%%%%%%%%%%%%%
\begin{figure} [H] 
\centering
\ \ \ \ \ \ \ \ \ \ \ \ \ \  {\includegraphics[width=0.96\textwidth]{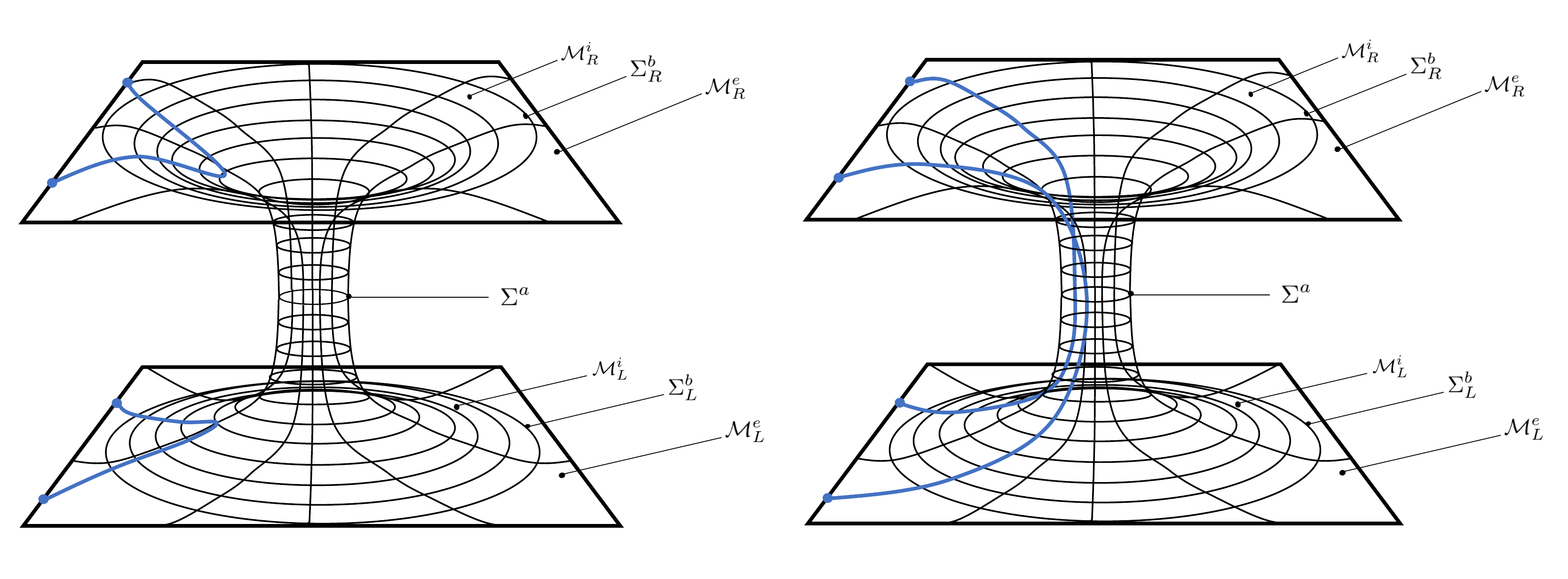}}
\vspace{-0.35cm}
\caption{{Sketch of the wormhole geometry and strings probing it.}} 
\lab{DibujoWormhole}
\end{figure}
%%%%%%%%%%%%%%%%%%%%%%%%%%%%%%%%
%%%%%%%%%%%%%%%%%%%%%%%%%%%%%%%%

Notice also that one can re-scale the coordinates and define the new dimensionless coordinates $\bar{x}^{\m}=(\bar{t},z,\vec{\bar{x}})$ for each patch, where $\bar{t} = {a}\,t/\ell^2 $, $z = {r}/{a} $, and $\bar{x} = {a}\,x/\ell $, and then write the metric as follows:
\be
ds^2 = \ell^2 \lp - \bar{f} \, d\bar{t}^2 + \bar{f}^{-1} \, dz^2 + z^2 \, d\vec{\bar{x}}^{\,2} \rp  \ ,
\ee
with
\be %\label{barf}
\bar{f}_i(z) = z^2 + 2\sqrt{3 - 2/z^2} \;, \;\qquad \bar{f}_e(z) = z^2 - 2\sqrt{\bar{\mu} - 2/z^2} \;  \,, 
\ee
having absorbed the dependence on $a$. In these coordinates, the throat of the wormhole is located at $z=1$. An adapted line element function defined by re-scaling $\bar{f}_i(z) \to N_i^2 \bar{f}_i(z)$ with $ N_i^2 = \bar{f}_e(z_b)/\bar{f}_i(z_b)$ and $z_b = \sqrt{\d}$, would be better suited to define a global time coordinate. This makes evident that there is a single relevant scale $\ell$ in the problem, being the radius of AdS$_5$.

%------------------------

\section{String dioptrics}

\subsection{The strings going through the thin-shell}

Let us first study the induced metric $h_{m n}$ on the timelike shell hypersurfaces; these are $\S^b_L$ and $\S^b_R$ for the right and left bubbles and $\S^a$ for the throat. Each of them is described by some intrinsic coordinates $\xi^m=(\xi^0,\xi^1,\xi^2, \xi^3)$ given by the embedding equations $x^{\m} = x^{\m}(\xi^m)$ on the hypersurface. The induced metric can be defined by starting with the vectors $e^{\m}_{m} = \p x^{\m}/\p \xi^{m}$, which are tangent to the curves of such hypersurface; more concretely, $h_{m n} =  e_m^{\m} e_n^{\n} \, G_{\m \n}$.
For later use, we define the projection tensor 
\be
h^{\m \n} \equiv  h^{m n} e^{\m}_{m} e^{\n}_{n} \ ,
\ee
useful to eliminate the normal components of an arbitrary tensor over the manifold. This is $G^{\m \n} = \, n^{\m} n^{\n} + h^{\m \n}$,
with $n^{\m}$ being the unitary normal vector to the shell hypersurface.
The line element of the induced metric over a shell of radius $r= \upsilon$ is given by
\be\label{inducedshellmetric}
dh_{\pm}^2 =  -f_{\pm}( \upsilon)\, dt_{\pm}^2 + \upsilon^2 d\vec{x}_{\pm} \,,
\ee
where $f_{\pm}( \upsilon)$ represent the behavior of the metric functions on each side of the juncture, i.e. $\upsilon =a$ for the throat, and $ \upsilon=b$ for the bubbles. Notice also that, the induced metric on the hypersurface of a shell at $r= \upsilon$ is unique, and the tangential coordinates at both sides are related by the first fundamental form which establishes the diagonal transformation
\be \label{tcoords}
{\L^{\m_+}}_{\m_-} = \mbox{diag} \lp \frac{\p t_+}{\p t_-}, \frac{\p r_+}{\p r_-} ; \frac{\p \vec{x}_+}{\p \vec{x}_-} \rp \,,
\ee 
with
\be \label{tcoords2}
\frac{\p t_+}{\p t_-} = \sqrt{\frac{f_{-}( \upsilon)}{f_{+}( \upsilon)}} \,, \qquad \frac{\p \vec{x}_+}{\p \vec{x}_-} = (1,1,1) \,,
\ee
which was obtained from (\ref{inducedshellmetric}). Summarizing, given a vector ${\bf U}$ with components $U^{\nu}$ in the bulk, its projection onto the shell hypersurface is 
\be
P({\bf U})^{\mu} =  h^{\mu \nu} \, U_{\nu} \ ;
\ee
then, if we considered different coordinates, e.g.,  $x^{\mu_+}$ and $x^{\mu_-}$, on the same hypersurface, we have that
\be \label{proy_2}
P({\bf U})^{\mu_+} = {\L^{\mu_+}}_{\mu_-} \, P({\bf U})^{\mu_-}  \,.
\ee

%-----

As mentioned in the Introduction, we are interested in probing the wormhole geometry with some particular configurations of fundamental strings that, in the context of the AdS/CFT correspondence, represent either an infinitely heavy meson, i.e. U-shaped strings with both of its endpoints on the same boundary, or an infinitely heavy isolated charge on each of the field theories, i.e. fundamental strings traversing the wormhole geometry from one asymptotic region to the other, to which we will refer to as LR-strings.
%represent infinitely heavy quarks (i.e. strings traversing the wormhole geometry from one asymptotic region to the other, to which we will refer to as LR-strings) or color neutral objects (i.e. U-shaped strings with both of its endpoints on the same boundary).
In order to study the string dynamics on this background, we consider the Nambu-Goto action given by
\begin{equation} \label{ng}
S= -\mathcal{T}_0\int{d\tau
d\sigma}\,\sqrt{-\det{g_{\alpha\beta}}} \,,
\end{equation}
where $g_{\alpha\beta}\equiv
G_{\mu\nu}\partial_{\alpha}X^{\mu}\partial_{\beta}X^{\nu}$ is the induced worldsheet metric, with $\sigma^{\alpha}\equiv (\tau, \sigma)$, $G_{\mu\nu}$ is the spacetime metric, and $\mathcal{T}_0= (2\pi{\alpha}')^{-1}$ is the tension of the fundamental string.  The projection of the worldsheet tangent vectors $\partial_{\alpha}X^{\mu} \equiv \p X^{\m} /\p\s^{\alpha}$ onto a hypersurface with induced metric $h_{m n}$, given by $h^{\m \n } \, \p_{\alpha} X_{\nu}$, are unique. This projection must be the same as seen from both sides of the hypersurface that defines the shell, so we have
\be
\lp h^{\m \n } \, \p_{\alpha} X_{\nu} \rp^+_- = 0 \,,
\ee
where $\pm$ here means the difference of the quantity evaluated at both sides of the shell. Since a priori different coordinates are to be used on each side, we can calculate the above expression using each coordinate system, as in (\ref{proy_2}), to get
\be \label{continuidad}
  h^{\m_+ \n_+ }  \p_{\alpha} X_{\nu+} =  {\L^{\m_+}}_{\m_-} \, h^{\m_- \n_- }  \p_{\alpha} X_{\nu-} \ ,
\ee
for a shell at $r= \upsilon$. We will use a static gauge to describe the map $X^{\m} (\t,\s) = ( t({\t}), r(\s), \vec{x}(\s))$ from the string worldsheet into spacetime --in each patch--, where we take $\vec{x}(\s)=(x(\s), 0, 0 )$. For a static string, the timelike and spacelike tangent vectors of the worldsheet are
\be
 \dot{X}^{\m} \equiv \p_{\t} X^{\m} (\s,\t) = ( \dot{t}, 0, 0, 0, 0 ) \, ,
\ee
\be
X'^{\m} \equiv \p_{\s} X^{\m} (\s,\t) =  ( 0, r', x', 0, 0 ) \, .
\ee
Evaluating (\ref{continuidad}) for $\s^{(1)} \equiv \t$ we obtain the condition at $r= \upsilon$ for timelike projections of the worldsheet, which in virtue of the diagonal metric tensors 
%, (\ref{inducedshellmetric}) and (\ref{tcoords}), 
is 
\be 
 \, {h^{t_+}}_{t_+} \dot{t}_+ = {\L^{t_+}}_{t_-}  \, {h^{t_-}}_{t_-}  \ , \dot{t}_-  \ ,
\ee
where the subscripts $\pm$ refer to coordinates on either one side or the other of the shell. In particular, for the wormhole geometry we would have
\be \label{t_punto}
\sqrt{f_+( \upsilon)} \, \dot{t}_+ = \sqrt{f_-( \upsilon)}\, \dot{t}_- \ .
\ee
Analogously, for $\s^{(2)} \equiv \s$, the spacelike projections of the worldsheet at $r= \upsilon$ are
\be 
 \, {h^{x_+}}_{x_+} {x'}_+ = {\L^{x_+}}_{x_-}  \, {h^{x_-}}_{x_-}  \, {x'}_- \ ,
 \ee
which for the wormhole geometry becomes simply
\be \label{x_prima}
 {x'}_+ = {x'}_-\, .
 \ee

%---------------------

\subsection{String refraction in the thin-shell wormhole}

In a generic static gauge, where $X^{\m} (\t,\s) = ( t({\t}), r(\s), x(\s), 0, 0 )$, the induced worldsheet metric is given by
\be
g_{\a \b} = \left(\begin{array}{cc}
-f(r) \, \dot{t}^2 & 0  \\
0 & \frac{r'^2}{f(r)} + r^2 \, x'^2
\end{array}\right)\ ,
\ee
and the corresponding Nambu-Goto action takes the form
\begin{flalign}
S  &=  - \mathcal{T}_0 \int d\tau d\s \sqrt{\dot{t}^2 \lp r'^2 + f(r) \, r^2 \, x'^2 \rp } \,,
\end{flalign}
while the canonical momentum densities of the string are defined as $\Pi^{\tau}_{\mu} \equiv - \mathcal{T}_0 \, \partial \sqrt{-g}/\p \dot{X}^{\mu}$, and $\Pi^{\sigma}_{\mu} \equiv - \mathcal{T}_0 \, \partial \sqrt{-g}/\partial X'^{\mu}$, with $\det{g_{\alpha\beta}} \equiv g$.
The string equations of motion supplemented with (\ref{tcoords})--(\ref{tcoords2}), imply that the tangential component of $\Pi^{\s}_{\mu}$ is continuous at its intersection with a shell at fixed $r(\s_{\upsilon})$, i.e. $\Pi^{\s}_x |^+_- =0$ in our case. Expressed in terms of different coordinate systems on each side, the latter condition reads 
\be \label{Px_cont}
\Pi^{\s}_{x_+} = {\L^{x_-}}_{x_+}  \Pi^{\s}_{x_-} \,,
\ee 
with ${\L^{x_-}}_{x_+} = 1 = ( {\L^{x_+}}_{x_-})^{-1}$ for the spatial coordinates of the shells in the wormhole. The action is time-independent and does not depend explicitly on the $x$--coordinate; the latter implies that $\Pi^{\s}_x$ is a conserved quantity, and it can be written as 
\be \label{constant}
C = - \frac{\Pi^{\s}_x}{\mathcal{T}_0} =  \frac{ |\dot{t}| f(r) \, r^2 \, x'}{ \sqrt{ r'^2 + f(r) \, r^2 \, x'^2 }} \ ,
\ee
with $C$ an integration constant. Applying equation (\ref{Px_cont}) at the throat, $r_{L}(\s_{a})=r_{R}(\s_{a})=a$, we have
\be
%C = 
\frac{  |\dot{t}_L|  f_i(a) \, a^2 \, x_L'}{ \sqrt{ r_L'^2 + f_i(a) \, a^2 \, x_L'^2 }}\Bigg|_{\s_{a}} =  \frac{   |\dot{t}_R| f_i(a) \, a^2 \, x_R'}{ \sqrt{ r_R'^2 + f_i(a) \, a^2 \, x_R'^2 }} \Bigg|_{\s_{a}} \ ,
\ee
where we have replaced $\mp$ subscripts by $L$ or $R$, and, using (\ref{t_punto}) and (\ref{x_prima}) at the throat, we obtain 
\be \label{ref_throat}
|r_L'| = |r_R'| \,.
\ee 
Analogously, situated at the position of the corresponding bubbles, $r_{L,R}(\s_b)=b$, on the Left or the Right hand side of the wormhole, we get
\be
%C = 
\frac{  |\dot{t_i}| f_i(b) \, b^2 \, x'_i}{ \sqrt{ r_i'^2 + f_i(b) \, b^2 \, x_i'^2 }}\Bigg|_{\s_b} =  \frac{  |\dot{t_e}| f_e(b) \, b^2 \, x'_e}{ \sqrt{ r_e'^2 + f_e(b) \, b^2 \, x_e'^2 }} \Bigg|_{\s_b}  \ ,
\ee
where we replaced $\mp$ by $i$ or $e$ subscripts, and, using (\ref{t_punto}) and (\ref{x_prima}), we obtain the refraction condition on the thin-shell bubbles
\be \label{r_prima}
\frac{r_e'}{\sqrt{f_e(b)} } = \frac{r_i'}{\sqrt{f_i(b)}} \,.
\ee
We note that the string refraction at the throat, (\ref{ref_throat}), which is trivial in virtue of the $\mathbb{Z}_2$-symmetry of the wormhole configuration, actually takes the sign adjustment $r_L' = -r_R'$ due to the orientation of the radial coordinates, both $\mathcal{M}_{L,R}^i$ are interiors, with $r$ pointing toward the corresponding boundary. On the other hand $x'(\s)$ is given by (\ref{constant}), and using the coordinate patch for each of the four regions in the wormhole spacetime, we have
\be \label{x_p_sigma}
x'(\sigma) = \left\{\begin{array}{cc}
\mp \frac{r'}{r\, \sqrt{f_e(r)}} \frac{C}{\sqrt{ \dot{t}^2 \,  r^2 \,  f_e(r) - C^2 }}\Big|_{L,e}  & \qquad  \mbox{for $\mathcal{M}^e_L$ }\\ \\
\mp \frac{r'}{r\, \sqrt{f_i(r)}} \frac{C}{\sqrt{ \dot{t}^2 \,  r^2 \,  f_i(r) - C^2 }}\Big|_{L,i}  & \qquad  \mbox{for $\mathcal{M}^i_L$ }\\ \\
\pm \frac{r'}{r\, \sqrt{f_i(r)}} \frac{C}{\sqrt{ \dot{t}^2 \,  r^2 \,  f_i(r) - C^2 }}\Big|_{R,i} & \qquad  \mbox{for $\mathcal{M}^i_R$ }\\ \\
\pm \frac{r'}{r\, \sqrt{f_e(r)}} \frac{C}{\sqrt{ \dot{t}^2 \,  r^2 \,  f_e(r) - C^2 }}\Big|_{R,e} & \qquad  \mbox{for $\mathcal{M}^e_R$ }
\end{array} \right.
\ee
which consistently satisfies (\ref{x_prima}). 
An explicit static gauge can be chosen for the time coordinates, namely $t_e(\tau)=\tau$ in the exterior regions $\mathcal{M}^e_{L,R}$ and $t_i(\tau)= \tau\, \Lambda^{t_i}_{\; t_e}|_{r=b} = \tau\, \sqrt{f_e(b)/f_i(b)}$ for the inner regions $\mathcal{M}^i_{L,R}$; these, in agreement with (\ref{t_punto}).

%-----------------

\section{Phase space}

We are interested in probing the wormhole geometry we constructed in section 2 with different static string configurations as those described in section 3. As explained before, these configurations are dual to either an isolated infinitely heavy quark in each theory (i.e. a fundamental LR-string with each of its endpoints on different boundaries) or to an infinitely heavy meson (i.e. a U-shaped string with both of its endpoints on the same boundary). We will study a particular scenario where a phase transition between these two configurations take place and try to shed some light on the dynamics of the two gauge theories. 

According to the standard recipe \cite{Malda2}, in the large-$N_c$, large $\lambda$ limit, the expectation value of a quark-antiquark pair tracing a rectangular Wilson loop where the temporal side is infinite, is dual to the on-shell Nambu-Goto action for a string worldsheet subject to the boundary condition of meeting the curve $\mathcal{C}$ at the boundary, i.e., 
\begin{equation}\label{wilsonnmalda}
\langle W(\mathcal{C})\rangle = e^{-i S(\mathcal{C})} \ .
\end{equation}
From (\ref{wilsonnmalda}) one can read off the energy of the corresponding quark-antiquark pair. This quantity is usually divergent, as the proper area of the string worldsheet is infinite; however, it can be regularized for example, by subtracting the proper area of two straight strings, cf. \cite{brandhuber, Rey}. Here, we will consider the regularized energy of the quark-antiquark pair defined by
\begin{equation}
\Delta E(L)=-\Delta S(L)/T \, ,
\end{equation}
where $T$ stands for the infinite time interval at the boundary, $\Delta E(L)=E-E_{ss}$, $\Delta S(L)=S-S_{ss}$, with $E_{ss}$ and $S_{ss}$ being the energy and on-shell action of a straight LR-string, respectively, and where $L$ is the spatial distance on the Wilson loop or, equivalently, the separation between the string endpoints. The latter is given by
\be \label{L}
L = \int_{-\frac{L}{2}}^{\frac{L}{2}} \, d(\ell x) \, , 
\ee
where we have incorporated the effective AdS radius $\ell$, in order to set the speed of light to 1 at the boundary theory.
In what follows, we will compute the energy $\Delta E$ from different string configurations refracted by the thin-shells of the wormhole. 
It is worth mentioning that the Euclidean version of this computation, corresponds to determining which string configuration has the lowest free energy $\mathcal{F}$, i.e. which string configuration is thermodynamically favored. 

Let us begin by consider a straight LR-string, this is a straight string which extends across the wormhole with its endpoints at different boundaries. The bending constant of this configuration is $C=0$, so the string pierces the shells perpendicularly, with $x'(\s)=0$, and there is no spatial refraction. The area density of the straight string is given by $|\dot{t}(\tau) \, r'(\s)|$, and the corresponding on-shell action takes the form,
\begin{flalign}
S_{ss}
 = - \mathcal{T}_0 \, 
\left(
2 \int_{i} d\tau |\dot{t}_i| \, d\s |r_i'| 
+
2 \int_{e} d\tau |\dot{t}_e| \, d\s |r_e'| 
\right) = -  \frac{\sqrt{\lambda } \, T}{\pi \ell^2}  
\left(
N_i  \int_{a}^{b} \, dr 
+
\int_{b}^{\infty} \, dr 
\right) \ ,
\end{flalign}
where the lapse factor $N_i \equiv (\p t_i/\p t_e)|_b$ suffices to adapt the rate of coordinate time of the inner regions with respect to that of the outer regions, and the exterior time coordinate is identified with the boundary time, $T_e = T$. For further convenience, hereafter, we will use a dimensionless normalized energy
\be \label{A}
\Delta \bar{E} \equiv -  \frac{ \Delta{S}/  T 
}{ a \, \mathcal{T}_0}  \ ,
\ee
with $a$ being the radius of the throat, and $\mathcal{T}_0 = \sqrt{\lambda}/(2\pi\, \ell^2)$. The normalized energy $\bar{E}_{ss}$ associated to the straight string configuration is then
\begin{flalign}
\bar{E}_{ss}
&= 2 \left( N_i  \int_{1}^{z_b} \, dz  +\int_{z_b}^{\infty} \, dz  \right) \ ,
\end{flalign}
with $z_b \equiv b/a \simeq 1.40$ being the position of the outer shells in terms of the variable $z=r/a$. The lapse factor is given by
\be \label{lapse} 
N_i \equiv  \frac{\p t_i}{\p t_e}\Big|_b %= \sqrt{\bar{f}_e(z_b)}/\sqrt{\bar{f}_i(z_b)} 
= \frac{\sqrt{\bar{f}_e(z_b)}}{\sqrt{\bar{f}_i(z_b)}} \simeq 0.20\,, 
\ee 
with
\be \label{barf} \bar{f}_i(z) \equiv z^2 + 2\sqrt{3 - 2/z^2} \;, \;\qquad \bar{f}_e(z) \equiv z^2 - 2\sqrt{\bar{\mu} - 2/z^2} \;  \,,  \ee
and $\bar{\m} \simeq 1.81$, as given in (\ref{M_b}). To analyze the behavior of the normalized dimensionless energy $\Delta{\bar{E}}$, we will use a normalized dimensionless quantity $\bar{L}$ defined as
\be \label{barL}
\bar{L} 
\equiv \frac{a \, L}{\ell^2}\,,
\ee
which represents the separation between the string endpoints.
In the following subsections, we will study the 
different possible configurations probing the wormhole spacetime and analyze the dynamics of the string dioptrics.

\subsection{U-shaped strings}

U-shaped strings have a turning point at $r_u=r(\s_u)$, which can be, in principle, either in an exterior or an interior region of the wormhole geometry. To evaluate the value of the bending constant $C$ for these configurations we have to take into account that
\be
\lim_{\s \to \s_u} x'(\s) = \mp \infty \ ,
\ee
and consequently, using (\ref{x_p_sigma}),
\be \label{constant_U}
C =  r_u \, \dot{t}  \sqrt{f(r_u)} = \left\{\begin{array}{cc} 
C_e = r_u \, \dot{t}_e \sqrt{f_e(r_u)}  %= r_u   \sqrt{f_e(r_u)} / \sqrt{f_e(b)}
 &\quad \mbox{if $r_u$ is in $\mathcal{M}^e$ } \\
C_i = r_u \, \dot{t}_i \sqrt{f_i(r_u)}  %= r_u  \sqrt{f_i(r_u)} / \sqrt{f_i(b)}
&\quad \mbox{if $r_u$ is in $\mathcal{M}^i$ }\\
\end{array} \right.
\ee 
where we have chosen $C$ to be positive. The value of such constant depends on the position of the turning point $r_u$ and determines how much the string bends. From inspection of the radicand given in (\ref{x_p_sigma}) we see that the U-shaped solutions must have $r\geq r_u$, this is because the product $\dot{t}^2 r^2 f(r)$ has a minimum at the throat and it grows with $r$, as given by the functions $f_i(r)$ and $f_e(r)$ in the intervals $a < r \leq b$ and $b \leq r$, respectively. The latter means that a U-shaped string with its turning point at one side of the wormhole must have both of its endpoints on the same boundary. 
We will eventually find three possible types of U-shaped string configurations. In what follows we will compute the separation between endpoints $\bar{L}$, given by 
\be \label{L_U}
\bar{L} =  \frac{2\,a}{\ell} \int^{{\infty}}_{\s_u}  d\s \, x'(\s) \,, 
\ee
with a parameterization such that $r(\s=\infty) = \infty$, and the regularized energy $\D\bar{E}$, obtained by subtracting $\bar{E}_{ss}$ to the corresponding normalized energy, written generically as
\be \label{energy}
\bar{E} = \frac{1}{a \, T 
} \int d\tau d\s |\dot{t}| \,\sqrt{r'^2 + f(r) \, r^2 \, x'^2} \,.
\ee
The values of $\bar{L}$ and $\bar{E}$ will permit us to compare between different string configurations.

\subsubsection{U-shaped strings in an exterior region}

For U-shaped strings with turning point $r_u$ in an exterior region of the wormhole $\mathcal{M}^e$ (left or right), the radial coordinate in the static gauge can be chosen to be $r(\s)=\s$. Using the expression for $x'(\s)$ given in (\ref{x_p_sigma}) and the bending constant $C_e$ as in (\ref{constant_U}), the separation between the two endpoints of the string is found to be
\be \label{bar_L}
\bar{L}  =   \frac{2 }{ z_u \, \sqrt{\bar{f}_e(z_u)}} \int^{\infty}_{z_u}  dz \, \frac{\bar{\b}_e}{\sqrt{1-\bar{\b}_e}} \,,
\ee
where $z_u =r_u/a$ can take values in the outer region, i.e. $z_u \in [ z_b ; \infty]$ with $ z_b \equiv b/a \simeq 1.40$, and where the auxiliary function is  
\be
\bar{\b}_e = \lp \frac{z_u}{z} \rp^2 \frac{\bar{f}_e(z_u)}{\bar{f}_e(z)} \,.
\ee
Then, the energy $\Delta\bar{E}(\bar{L})$ for a quark-antiquark pair dual to the U-shaped string with its turning point at $r_u$ in an exterior region of the wormhole geometry is given by
\be \label{bar_A}
\Delta\bar{E}  = 2 \left( \int_{z_u}^{\infty}  dz \, \frac{1-\sqrt{1 - \bar{\b}_e}}{\sqrt{1 - \bar{\b}_e}} 
- (z_u - z_b)- \,N_i (z_b - 1) \right) \,.
\ee
This configuration is similar to those studied in black hole backgrounds, however, here the outer region corresponds to a metric with supracritical value of charge $Q$. Notice that, for small separations, corresponding to turning points $z_u \gg z_b$, we have $\bar{\b_e} \sim (z_u/z)^{4}$ and, as expected, we recovered the Coulomb behavior $\Delta\bar{E} \sim -1/ \bar{L}$ dictated by the conformal invariance at the UV region of the dual field theory.
The energy $\Delta\bar{E}$ as a function of $\bar{L}$ for a quark-antiquark pair described by these U-shaped strings whose turning point is in an exterior region is plotted with a blue curve in figure \ref{AvsL_U}. The expressions for these quantities, (\ref{bar_A}) and (\ref{bar_L}), are independent of the physical scales in virtue of the auxiliary normalized bar functions. The maximum endpoints separation for these configurations is found to be $\bar{L}^{bu} \simeq 1.04$, corresponding to a U-shaped string whose turning point barely stays inside the outer region of the geometry, i.e. just above the bubble.

%---------------------------------

\subsubsection{Refracted U-shaped strings: turning point in the interior region}

If the turning point $r_u$ is in an interior region $\mathcal{M}^i$ (left or right), the string dioptrics is characterized by the refraction law given by (\ref{r_prima}) at the bubble, and an explicit radial coordinate parameterization in the static gauge can be written as
\be \label{r_sigma_U}
r(\s) = \left\{\begin{array}{lll}
- (b-r_u) \frac{\sqrt{f_e(b)}}{\sqrt{f_i(b)}} \; (\s + 1) + b \qquad  \qquad & \s \in(-\infty; -1]  \mbox{, for $\mathcal{M}^e_{L,R}$ }\\
-(b-r_u)  \;\s + r_u \qquad  & \s \in[-1; \sigma_u=0]  \mbox{, for $\mathcal{M}^i_{L,R}$ }\\
(b-r_u)  \;\s + r_u \qquad &  \s \in[\sigma_u=0; 1] \mbox{, for $\mathcal{M}^i_{L,R}$ }\\
(b-r_u) \frac{\sqrt{f_e(b)}}{\sqrt{f_i(b)}} \; (\s - 1) + b \; \qquad  & \s \in[1; +\infty)  \mbox{, for $\mathcal{M}^e_{L,R}$ }
\end{array}\right.
\ee
Using the corresponding expression for $x'(\s)$ in each region, as in (\ref{x_p_sigma}), and the bending constant $C_i$ from (\ref{constant_U}), the distance $\bar{L}$ between endpoints of the refracted U-shaped string is 
\begin{flalign} \label{barL_Ui}
\bar{L} =  \frac{2}{z_u \,N_i\,  \sqrt{\bar{f}_i(z_u)}} \left(  N_i \int^{z_b}_{z_u}  dz \; \frac{\bar{\b}_i}{\sqrt{1-\bar{\b}_i}} +  \int^{\infty}_{z_b}  dz \; \frac{\bar{\b}_{ie}}{\sqrt{1-\bar{\b}_{ie}}} \right) \ ,
\end{flalign}
with the auxiliary functions defined as
\be
\bar{\b}_i = \lp \frac{z_u}{z} \rp^2 \frac{\bar{f}_i(z_u)}{\bar{f}_i(z)} \,, \quad
\bar{\b}_{ie} = \lp \frac{z_u}{z} \rp^2 \frac{\bar{f}_i(z_u) }{\bar{f}_e(z)} \, N_i^{2} \,,
\ee
and $z_u = r_u/a$ being the turning points in the interior region, taking values $z_u \in [1 ; z_b]$. Notice that the presence of the lapse $N_i$ in the above expressions makes clear the refractions when going through the shell hypersurface at $r=b$. The corresponding energy $\Delta\bar{E}$ for these string configuration is
\begin{flalign} \label{barA_Ui}
\Delta\bar{E} = 2 \lp 
N_i \int_{z_u}^{z_b} dz \; \frac{1-\sqrt{1 - \bar{\b}_i}}{\sqrt{1 - \bar{\b}_i}} +
\int_{z_b}^{\infty} dz \; \frac{1-\sqrt{1 - \bar{\b}_{ie}}}{\sqrt{1 - \bar{\b}_{ie}}}
- N_i (z_u-1)  \rp \,.
\end{flalign} 

\begin{figure}  [H] 
\centering
  \begin{minipage}{0.5\textwidth}
    \centering	
    \includegraphics[width=0.94\textwidth]{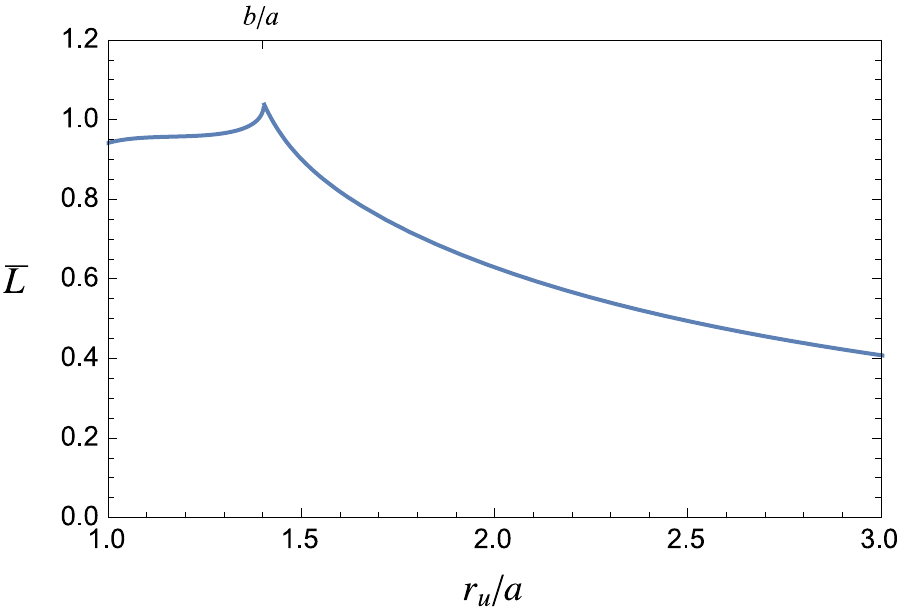}
  \label{}
  \end{minipage}%
  \begin{minipage}{0.5\textwidth}
    \centering
    \includegraphics[width=0.97\textwidth]{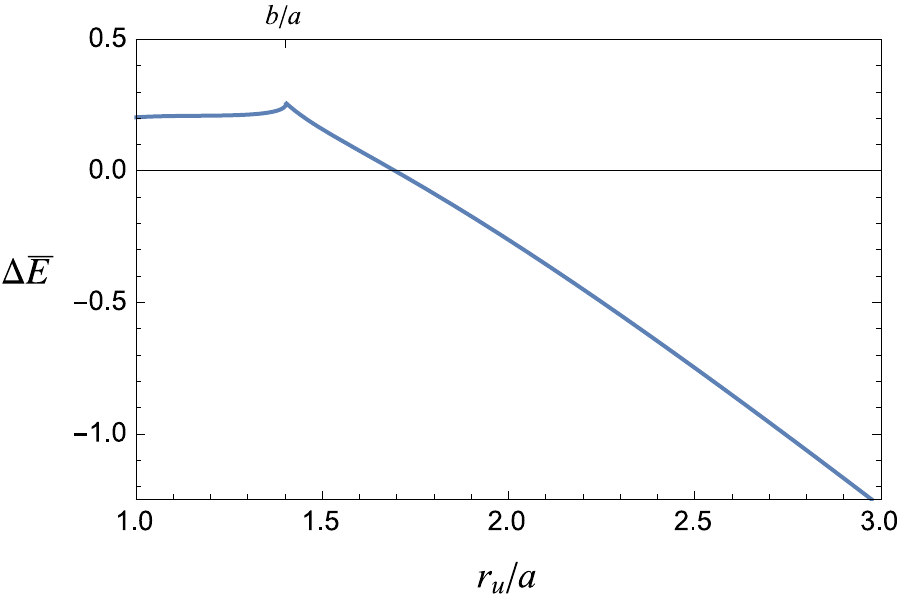}
      \label{}
  \end{minipage}
  \vspace{-0.4cm}
  \caption{$\bar{L}$ and $\D\bar{E}$ as functions of the turning point $r_u/a$.}
  \label{LyAvstp}
\end{figure}

Figure \ref{LyAvstp} shows the distance $\bar{L}$ and the energy $\Delta\bar{E}$ as function of the ratio $r_u/a$ for U-shaped string configurations with turning point in the outer and inner bulk regions. Notice that both the energy $\Delta\bar{E}$ and the separation $\bar{L}$ have a maximum if the turning point is at the position of the bubble's shell $b \simeq 1.40 \,a$. For refracted configurations, the separation $\bar{L}$ decreases as the turning point is moved in the inner region deeper into the wormhole, and the minimum endpoints separation of refracted U-shaped strings is obtained for a turning point just above the throat, this value is $\bar{L}^{th} \simeq 0.94$. 
A numerical plot of $\Delta\bar{E}$ as a function of $\bar{L}$ corresponding to U-shaped strings with turning point in the interior region is shown in figure \ref{AvsL_U} with a light blue curve. The segment between $\bar{L}^{th}$ and $\bar{L}^{bu}$ depicted in figure \ref{AvsL_U} resembles the fishtail tip featured for string configurations in black hole backgrounds: The refracted configurations, which extend deeper into the bulk, are energetically disfavored with respect to exterior U-shaped strings that remain outside the bubble.

%--------------------------------

\subsubsection{U-shaped strings stretched at the throat}

To complete the configuration space of U-shaped strings in the wormhole spacetime, we have to consider a third type of solutions: These correspond to configurations which dip into the inner region and have a finite length segment lying down on the throat at a fixed radius. Before constructing these solutions, we will first show the existence of another kind of configuration (not U-shaped) which corresponds to a string completely stretched at fixed $r=a$. We recall that the wormhole spacetime has reflection symmetry around the throat's shell and that gravity pulls towards this hypersurface located at $r=a$, which thus become an accumulation surface. If it were the case of a smooth geometry around the throat, when posing the equations of motion we would immediately find that there exists a static solution corresponding to the aforementioned string at fixed $r=a$, i.e. at the surface where the space opens up. The same happens in our case, in which the space opens up at the infinitely thin throat's shell where such a configuration also appears. To bear out this, we can adopt a new radial coordinate ${y}$, which intersects the throat hypersurface at ${y}=0$, and consider the metric around this position, namely
\be
ds^2 = - V({y}) \,dt^2 + V^{-1}({y})\, d{y}^2 + r^2({y}) \,d\vec{x}\cdot d\vec{x}  \, ,
\ee
with $r({y}) =  ({y}-a)  \, \Theta({y}/a) + (a-{y}) \, \Theta(-{y}/a) $, where $\Theta $ is the Heaviside step function, and $V({y}) = f_i(r({y}))$. Now, let us see that the equations of motion for a string at fixed ${y}$ are satisfied at ${y}= 0$. For concreteness, let us consider the string description with some static gauge in which the coordinates are $X^{\m} (\t,\s) = ( t({\t}), {y}(\s), x(\s), 0, 0 )$, so the area density and momenta are given by
\be \label{newconditions1}
\sqrt{-{g}} =  \dot{t} \sqrt{  {y}'^2 + V \, r^2 \, x'^2  } \, ,
\ee
and
\be\label{newconditions2}
\Pi^{\tau}_{t}= - \mathcal{T}_0\sqrt{  {y}'^2 + V \, r^2 \, x'^2  }
\qquad
\Pi^{\s}_{{y}} =  - \mathcal{T}_0 \frac{ \dot{t}\, {y}' }{ \sqrt{ {y}'^2 + V \, r^2 \, x'^2 }} 
\qquad
\Pi^{\s}_x  = - \mathcal{T}_0 \frac{ \dot{t} \, V \, r^2 \, x'}{ \sqrt{ {y}'^2 + V \, r^2 \, x'^2 }} \ .
\ee
To describe a string at a fixed radial coordinate ${y}$, we use a parameterization where $\s=x$. When plugging $x'=1$ and ${y}'=0$ in (\ref{newconditions1}) and (\ref{newconditions2}), we see that the equations of motion for such configuration are satisfied if $\p \sqrt{-{g}}/\p {y} = 0$.
The latter reduces to the condition $r'({y})= 0$, and considering the step functions involved in the metric coefficients, we have that it is achieved at the minimum radius $r=a$, i.e at the throat.

We are not specifically interested in the solution described above; instead, we will consider it just as a piece of the string solution, to complete the U-shaped string configuration: We are looking for solutions with endpoints at the boundary; therefore, we take half of a U-shaped string, we cut it at its turning point, which lies down tangentially just over $r=a$, and then attached to it one end of a finite-length segment lying on the throat, with the momenta being continuous. The same procedure is performed in the other end, so obtaining the U-shaped solution: a string with endpoints at one boundary, refracted at the bubble's shell, and stretched tangentially to the throat. For the stretched part of the string, parameterized with $\s = x$, we have  ${y}'=0$ and, consequently, the momenta conservation and the density continuity condition at $r=a$ are satisfied. 
In particular, for the stretched part localized at $r=a$ the area density is 
\be
\sqrt{-{g}}|_{r=a} =  a\, \dot{t}_i \,\sqrt{  f_i(a) } \,.
\ee
Note that this segment along the $x$ direction %, at the throat's position, 
can be stretched an arbitrary length $x_{str}$. In this way, the endpoints separation of a U-shaped string which has a segment $x_{str}$ lying stretched at the throat's position is
\be \label{barL_s}
\bar{L} = \frac{a}{\ell} \left( \int_{{-\infty}}^{\s_{a}}  d\s \, x'(\s) +
x_{str} 
+
\int^{{\infty}}_{\s_{a}}  d\s \, x'(\s)  \right)
=  \bar{L}^{th} +  \bar{L}^{str} \ ,
\ee
where we have named $\bar{L}^{str} \equiv a \,x_{str}/\ell$ in the last step, and 
\be
\bar{L}^{th}  =  \frac{2}{ \sqrt{3}} \left(  \int^{z_b}_{1}  dz \; \frac{\bar{\b}_i}{\sqrt{1-\bar{\b}_i}}\bigg|_{z_u=1} +  \frac{1}{N_i} \int^{\infty}_{z_b}  dz \; \frac{\bar{\b}_{ie}}{\sqrt{1-\bar{\b}_{ie}}}\bigg|_{z_u=1}
 \right) \ ,
\ee
is equivalent to the separation of a refracted U-shaped string with turning point just over the throat. %$\bar{f}_i(1)=3$
The energy $\Delta\bar{E}$ associated to the stretched U-shaped string configuration is then
\be \label{E_strU}
\Delta\bar{E}   = \Delta\bar{E}^{th} +  \bar{E}^{str} \ ,
\ee
with
\be \label{E_str}
%\D \epsilon 
\bar{E}^{str} = \frac{1}{a \, T %\D t
} \, \int d\tau  \, d\s \sqrt{-{g}}|_{r=a} %= \sqrt{3} \, N_i \, \D \bar{L} 
= \sqrt{3} \, N_i \, \lp \bar{L} - \bar{L}^{th} \rp \ ,
\ee
where we used that $\sqrt{f_i(a)} = a\sqrt{3}/\ell$, and with  
\be
\Delta\bar{E}^{th}  
= 2 \lp 
N_i \int_{1}^{z_b} dz \; \frac{1-\sqrt{1 - \bar{\b}_i}}{\sqrt{1 - \bar{\b}_i}}\Bigg|_{z_u=1} +
\int_{z_b}^{\infty} dz \; \frac{1-\sqrt{1 - \bar{\b}_{ie}}}{\sqrt{1 - \bar{\b}_{ie}}}\Bigg|_{z_u=1} 
 \rp \ ,
\ee
equivalent to the energy found for a refracted U-shaped string which almost touches the throat.

\begin{figure}  [H] 
\centering
  \begin{minipage}{0.49\textwidth}
    \centering	
    \includegraphics[width=.99\textwidth]{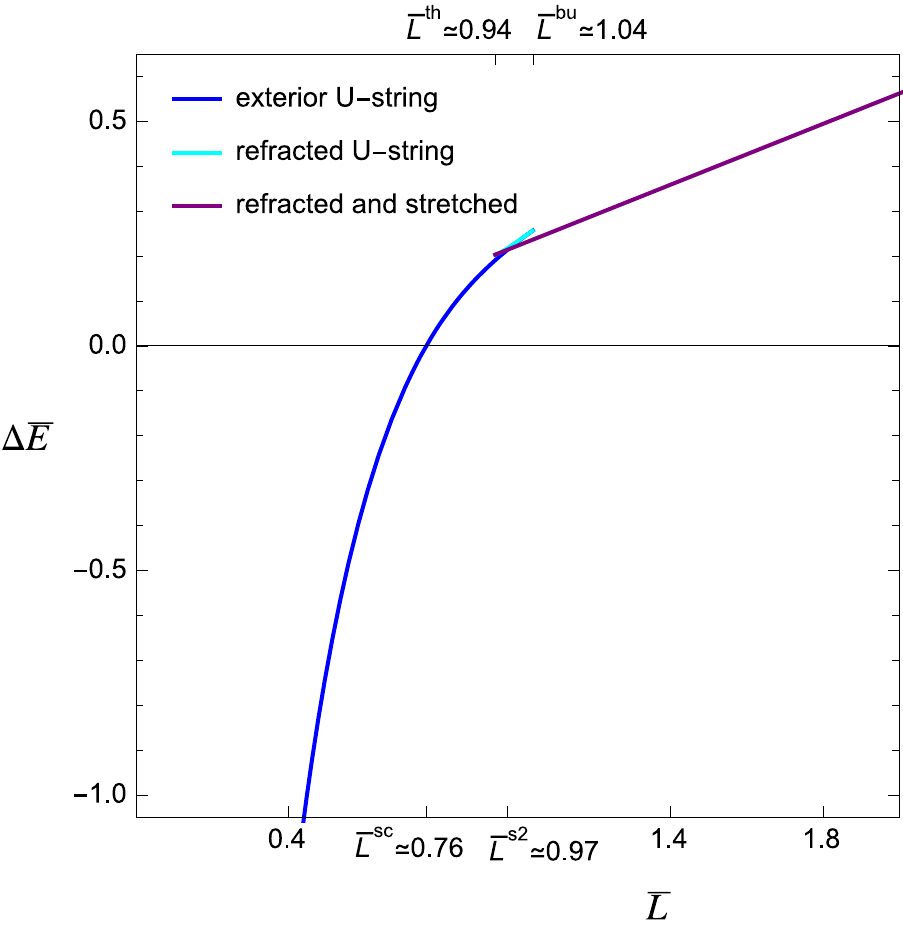}
    %\\{\footnotesize (a) $\bar{L}$ vs $\r_u/\r_a$.} 
%  \vspace{0.5cm}
  \label{EvsL_U_g}
  \end{minipage}%
 \hspace{1mm}
  \begin{minipage}{0.49\textwidth}
    \centering
    \includegraphics[width=0.99\textwidth]{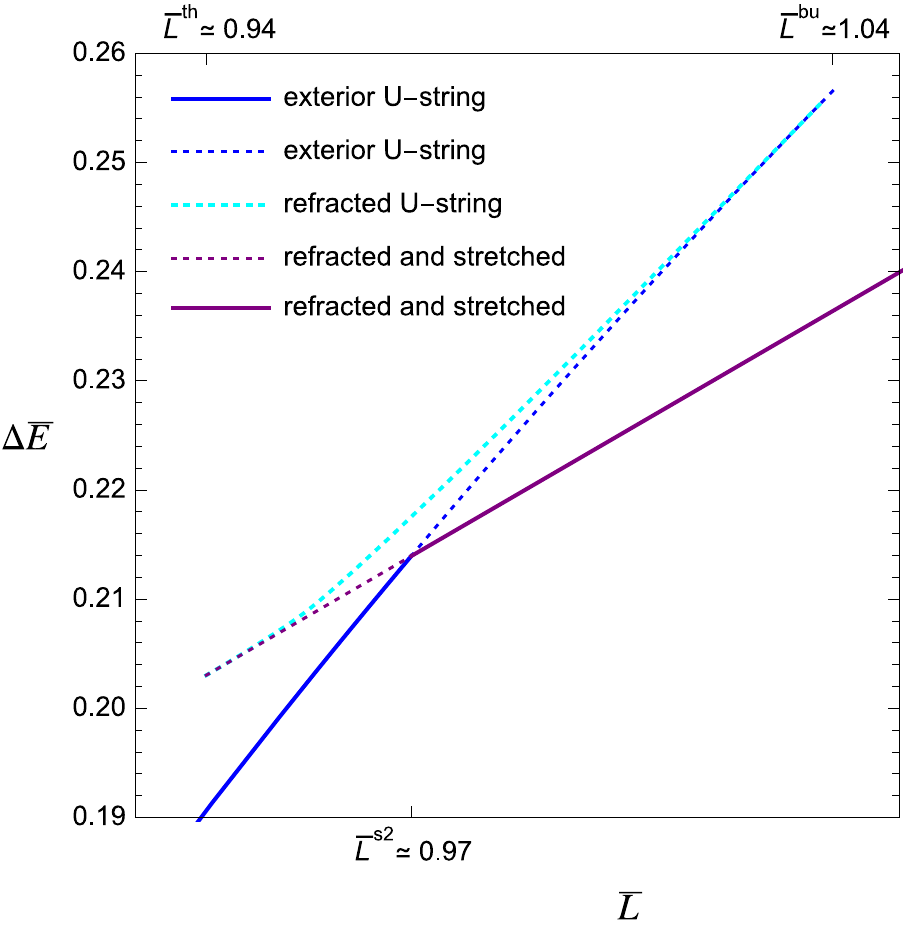}
    %\\    {\footnotesize (b) $\D\bar{E}$ vs $\r_u/\r_a$.} 
      \label{AvsL_U_detalle}
  \end{minipage}
  \vspace{-0.4cm}
  \caption{The energy $\D\bar{E}$ of a quark-antiquark pair as a function of its separation $\bar{L}$. The dual portrayal corresponds to a U-shaped string with both of its endpoints on the same boundary in the wormhole geometry.}
  \label{AvsL_U}
\end{figure}

We emphasize that these stretched solutions are not an artifact of the thin-shell throat; as we will see in section \ref{other_backgrounds}, analogous configurations appear in smooth wormholes geometries as well.
The energy $\Delta\bar{E}$ as a function of the endpoints separation $\bar{L}$ for all physically allowed U-shaped configurations, including those analyzed in previous sections, is shown in figure \ref{AvsL_U}: exterior U-shaped strings are plotted in blue, refracted U-shaped strings are represented in light blue, and stretched U-strings are depicted in purple. The figure on the right is a zoomed version of the plot on the left for the region where the different configurations share similar separation distance. Disfavored configurations are represented with dotted lines; among the U-shaped configurations, strings in the exterior region are energetically favorable for separations $\bar{L} \leq \bar{L}^{s2} \simeq 0.97$, and by increasing the separation from there on, the stretched U-shaped strings have the lowest area.
Nevertheless, the straight string used in the subtraction regularization is the preferred configuration for $\bar{L}  \geq \bar{L}^{sc}\simeq 0.76$, as can be seen from the intersection with the horizontal axis in the figure on the left. Despite the fact that the stretched U-shaped strings are energetically disfavored when compared with those going through the throat, notice that for such solutions, which explore the deep IR, the energy $\Delta\bar{E}$ increases linearly as a function of the endpoint separation $\bar{L}$. Specifically, this linearity is characterized by an effective string tension given by $ \sqrt{3}\, N_i \, \mathcal{T}_0\, a^2/\ell^2$, as obtained in (\ref{E_strU}) and (\ref{E_str}), after restoring dimensionful quantities.

%------------------------

\subsection{LR-strings}

When studying strings that extend from one side to the other of the wormhole geometry, we consider two different configurations: There are extended strings which traverse the wormhole piercing the throat's surface and, on the other hand, in the same fashion as the U-shaped strings discussed above, there are extended strings which traverse the wormhole but stretched a finite length parallel to the thin throat. We begin by characterizing the former case which can be parameterized in the static gauge as
\be \label{r_sigma}
r(\s) = \left\{\begin{array}{cc}
- (b-a) \frac{\sqrt{f_e(b)}}{\sqrt{f_i(b)}} \; (\s + 1) + b & \qquad  \s \in(-\infty; -1]  \mbox{, for $\mathcal{M}^e_L$ }\\
-(b-a)  \;\s + a &\qquad  \mbox{$\s \in[-1; \s_{a}$$=$$0]$, for $\mathcal{M}^i_L$ }\\
(b-a)  \;\s + a &\qquad \mbox{$\s \in[$$\s_{a}$$=$$0$$; 1]$, for $\mathcal{M}^i_R$ }\\
(b-a) \frac{\sqrt{f_e(b)}}{\sqrt{f_i(b)}} \; (\s - 1) + b \; & \qquad  \s \in[1; +\infty)  \mbox{, for $\mathcal{M}^e_R$ }
\end{array}\right.
\ee
These string configurations are refracted once in each bubble at radial coordinate, $r_{L,R}= b$. The separation between the endpoints, as measured in the $x$ direction of the Left and Right boundaries, will be denoted as $L_{LR}$, and it is calculated analogously as for the U-shaped strings. The normalized separation for these type of configurations is 
\be \label{barL_LR}
\bar{L}_{LR} \equiv \frac{a}{\ell} \left( \int_{{-\infty}}^{\s^-_{a}}  d\s \, x'(\s) + \int^{{+\infty}}_{\s^+_{a}}  d\s \, x'(\s)  \right) \,,
\ee
from which $L_{LR} =  \ell^2\, \bar{L}_{LR} /a$ is obtained. Explicitly, this yields
\begin{flalign} \label{barL_LR}
\bar{L}_{LR} =   \frac{ 2 \,\bar{C}}{ \sqrt{3}} \left(  \int^{z_b}_{1}  dz \; \frac{\bar{\varphi}_i}{\sqrt{1- \bar{C}^2 \, \bar{\varphi}_i}} +  \frac{1}{N_i} \int^{\infty}_{z_b}  dz \; \frac{\bar{\varphi}_{ie}}{\sqrt{1-\bar{C}^2 \,\bar{\varphi}_{ie}}} \right) \,,
\end{flalign}
where $z_b = b/a$, we have used that $\bar{f}_i(1) = 3$, and the auxiliary functions defined as
\be
\bar{\varphi}_i 
= \lp \frac{1}{z} \rp^2 \frac{3}{\bar{f}_i(z)} \,, \qquad
\bar{\varphi}_{ie} 
= \lp \frac{1}{z} \rp^2 \frac{3\, N_i^2}{\bar{f}_e(z)} \,.
\ee
The dimensionless constant $\bar{C}$ that appears in (\ref{barL_LR}) is the normalized positive integration constant $C$, as in (\ref{constant}). This parameter specifies the  configurations indicating how much the string bends. For configurations going through the throat, the possible values of this constant are read from the radicands in (\ref{barL_LR}), yielding $\bar{C} \in [0;1]$. Note that $\bar{C} = 0$ corresponds to the straight string, while $\bar{C} = 1$ produces the maximum bending which corresponds to a string with an inflection point at the throat (the string intersects the throat tangentially). Due to reflection symmetry, the $\bar{C} = 1$ configuration has the same length as the refracted U-shaped string whose turning point barely touches the throat of the wormhole tangentially. The latter can be checked analytically by evaluating (\ref{barL_Ui}) at $z_u \equiv r_u/a = 1$. The energy $\Delta\bar{E}$ obtained from this extended strings characterized by $\bar{C}$ is given as
\begin{flalign} \label{barA_LR}
\Delta\bar{E} = 2 \lp 
N_i \int_{1}^{z_b} dz \frac{1-\sqrt{1 - \bar{C}^2 \, \bar{\varphi}_i}}{\sqrt{1 - \bar{C}^2 \,  \bar{\varphi}_i}} +
\int_{z_b}^{\infty} dz \frac{1-\sqrt{1 - \bar{C}^2 \,  \bar{\varphi}_{ie}}}{\sqrt{1 - \bar{C}^2 \,  \bar{\varphi}_{ie}}} 
\rp \,.
\end{flalign} 
We can analyze the small endpoints separation behavior of a quark-antiquark pair in different boundaries by expanding the above expressions with $\bar{C} \ll 1$, this is
\be \label{barA_small_C}
\Delta\bar{E} = \bar{C}^2 \lp 
N_i \int_{1}^{z_b} dz \, \bar{\varphi}_i +
\int_{z_b}^{\infty} dz  \, \bar{\varphi}_{ie} 
\rp +
\mathcal{O} (\bar{C}^3) \,,
\ee 
\be \label{barL_small_C}
 \bar{L}_{LR} =  \bar{C} \, \frac{2}{\sqrt{3} \, N_i } \lp 
N_i \int_{1}^{z_b} dz \, \bar{\varphi}_i +
\int_{z_b}^{\infty} dz  \, \bar{\varphi}_{ie} 
\rp +
\mathcal{O} (\bar{C}^3) \,.
\ee
Combining the latter expansions we obtain
\begin{flalign} \label{A_L2}
\Delta\bar{E} \simeq \frac{1}{2} \, \bar{K} \, \bar{L}_{LR}^2  \,,
\end{flalign} 
with 
\be
\bar{K} = \frac{1}{2} \lbr
N_i \int_{1}^{z_b}  \frac{dz}{z^2 \, N_i^2 \, \bar{f}_i(z)} +
\int_{z_b}^{\infty} \frac{dz }{z^2 \, \bar{f}_e(z)} 
\rbr^{-1} \simeq 0.88 \,,
\ee
which exhibits a spring-like interaction for a pair of charges in different boundaries, as opposed to the Coulomb potential found for small separations of the quark-antiquark pair in the same gauge theory. When restoring dimensionful quantities, we have that for a pair separated a distance $L_{LR} \ll \ell^2/a$, the energy $\D E = \frac{1}{2} \, K \, L_{LR}^2$ is given by a spring-like effective constant $K = \bar{K} \, \mathcal{T}_0 \, a^3/\ell^4 \sim \sqrt{\lambda} \, a^3/\ell^6$. The energy increases with smaller $\ell$ and with greater $a$, consistently with bringing the charges closer to each other.

%------------

\begin{figure}  [H] 
\centering \begin{minipage}{0.45\textwidth}
\centering	\includegraphics[width=1\textwidth]{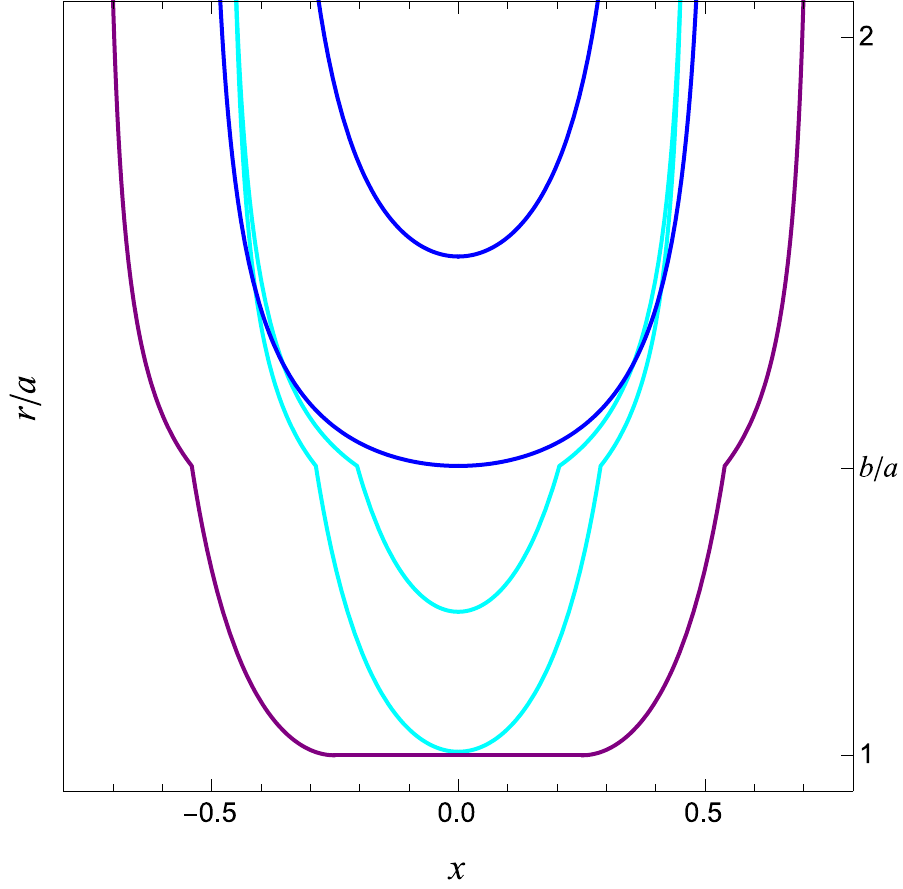} 
\\  
    {\footnotesize (a) U-shaped strings.} 
  \end{minipage}
\hspace{2mm}
  \vspace{-0.2cm}
  \begin{minipage}{0.45\textwidth}
    \centering    \includegraphics[width=1\textwidth]{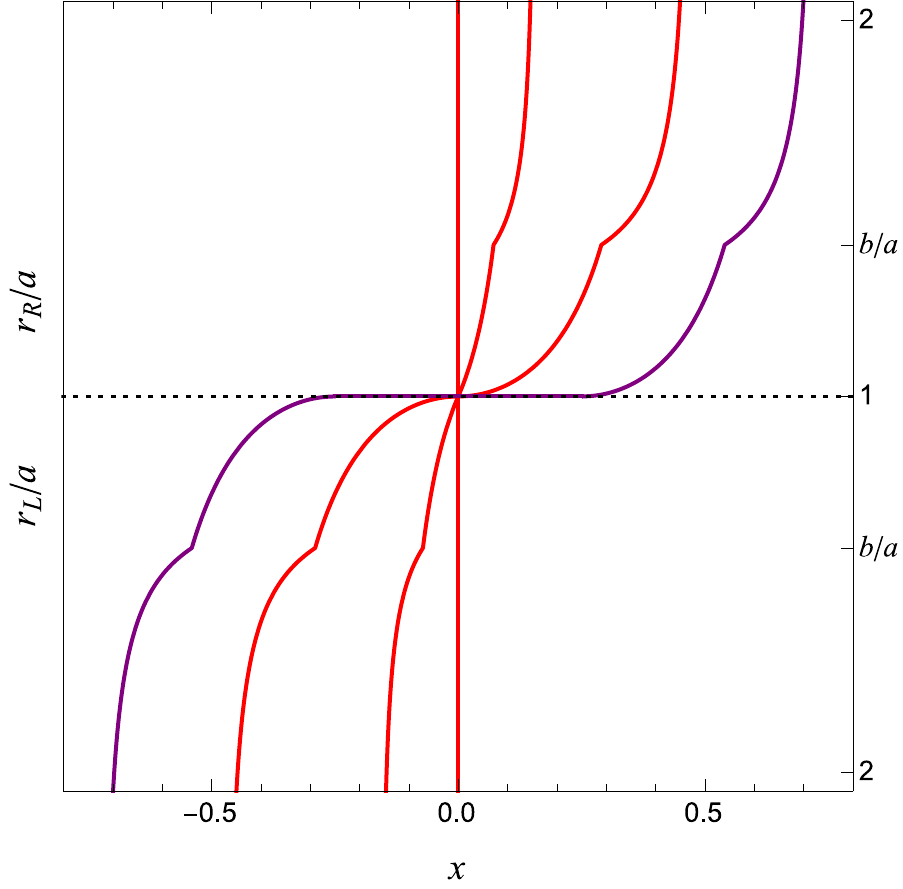}\\
 %   \vspace{-0.25cm}
    {\footnotesize (b) LR-strings.%with $\bar{C}= \{0, 0.5, 1$\},
} 
        \end{minipage}
          \caption{String dioptrics in the thin-shell wormhole.}
  \label{config}
\end{figure}

The remaining physically allowed configurations correspond to extended LR-strings stretching along the thin throat of the wormhole. To obtain such configurations, we pick half of a U-shaped string from each side of the wormhole with $\bar{C}=1$ (with the correct orientation), and apply the same procedure described for U-shaped strings to join them with a finite-length stretched string segment at $r=a$. Then, in virtue of the symmetry across the throat, the separation and associated energy are the same as the ones calculated previously for the stretched U-shaped strings; these are given in (\ref{barL_s}) and (\ref{E_strU}). In this way, stretched LR-configurations also show the linear dependence
\be
\Delta\bar{E} =  \sqrt{3} \, N_i\,  ( \bar{L}_{LR} - \bar{L}^{th} ) +\Delta\bar{E}^{th} \ ,
\ee
for $\bar{L}_{LR} \geq \bar{L}^{th}$, with dimensionful effective string tension given by $\sqrt{3}\,N_i\,\mathcal{T}_0\, a^2/\ell^2 \sim \sqrt{\l}\, a^2/\ell^4$. 
The dioptrics of strings configurations are displayed in figure \ref{config}, where we represented some examples in the ($r/a$,$x$)-plane. Figure \ref{config}(a) shows exterior U-shaped strings (in blue), refracted U-shaped strings (in light blue) and stretched U-shaped strings (in purple). Figure \ref{config}(b) shows extended LR-strings piercing the throat with bending constants $\bar{C}=0, 0.5, 1$ (in red), and a stretched LR-string (in purple).

The results involving the complete space of solutions are summarized in figure \ref{EvL} with the plot of $\D\bar{E}$ as a function of the endpoints separation $\bar{L}$: exterior and refracted U-shaped configurations are represented in blue and light blue, respectively; LR-string configurations piercing the throat are in red, and stretched configurations are in purple.
We observe there is a critical length $\bar{L}^{sc}$ for these configurations where a transition takes place: For separations $\bar{L}< \bar{L}^{sc}$, quark-antiquark pairs on the same boundary theory are energetically favorable, yielding U-shaped string configurations. If a second pair is present on the other boundary, charges in the same gauge theory can be screened by transitioning to LR-configurations connecting different boundaries. In other words, here we have managed to reproduce the same qualitative features observed in \cite{Vazquez1, Vazquez2}, but for the case of thin-shell wormholes with flat constant-$r$ hypersurfaces, namely for the case in which the gauge theory is formulated on $\mathbb{R}^{1,3}$.
 
\begin{figure} [H] 
\centering
{\includegraphics[width=0.49\textwidth]{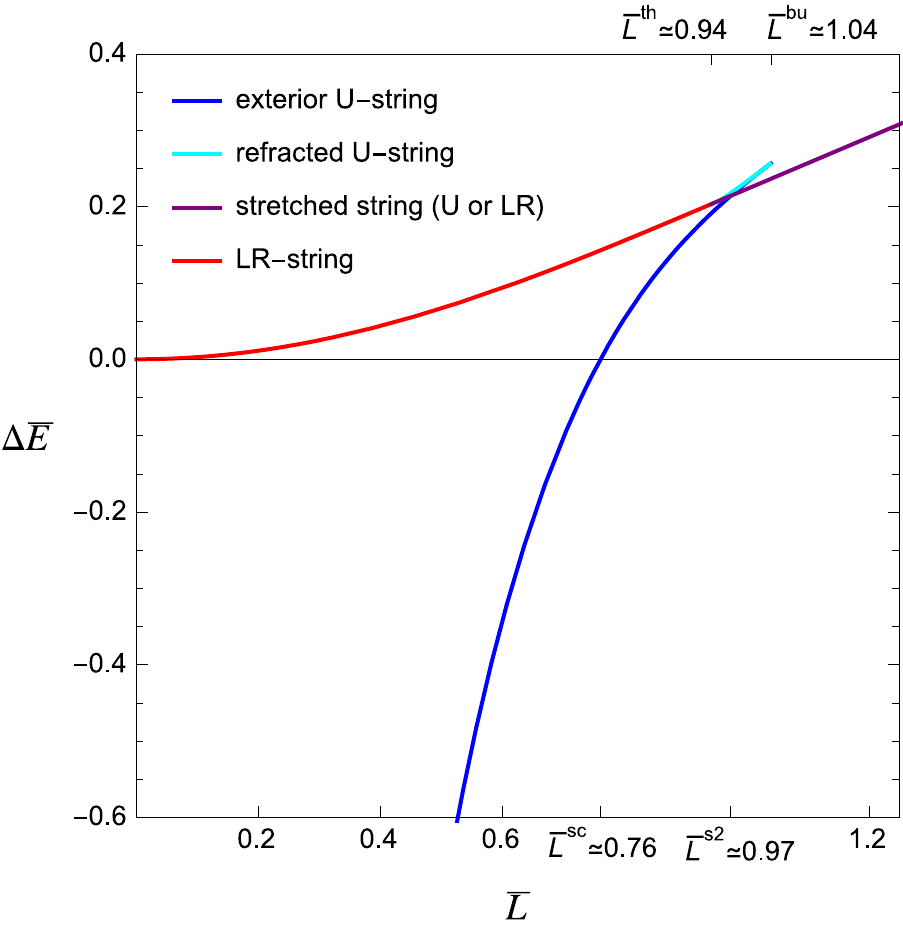}}
\vspace{-0.35cm}
\caption{The energy $\D\bar{E}$ of a quark-antiquark pair as a function of its separation $\bar{L}$. Configurations within the same boundary, dual to U-shaped strings, are plotted in blue, light blue and purple. Left-Right configurations, dual to strings which traverse the throat, are plotted in red and purple.} 
\lab{EvL}
\end{figure}

%-----------------

\section{Smooth wormholes}
\label{other_backgrounds}

Before we conclude, let us compare our results with those obtained in \cite{Vazquez2} where they considered string probes in smooth wormholes that connect two locally AdS$_5$ asymptotic regions. Such geometries were obtained in \cite{Julio} and are also solutions of 5D EGB theory at the CS point; the metric is given by
\be \label{DOT_metric}
ds^2 = \ell^2 \lp -\cosh^2{(\r-\r_0)} \, d{t}^2+ d\r^2 + \cosh^2(\r) \,d\S_3^2 \rp \ ,
\ee
where $d\Sigma_3^2$ now corresponds to a 3-dimensional space of negative constant curvature, locally equivalent to $\Sigma_3 = S^1 \times \mathbb{H}_2$, $\rho \in \mathbb{R}$, and where $\rho_0$ is an arbitrary constant that parameterizes the asymmetry of the wormhole: the configuration is $\mathbb{Z}_2$-symmetric for $\rho_0=0$.

\begin{figure} [H] 
\centering
{\includegraphics[width=0.49\textwidth]{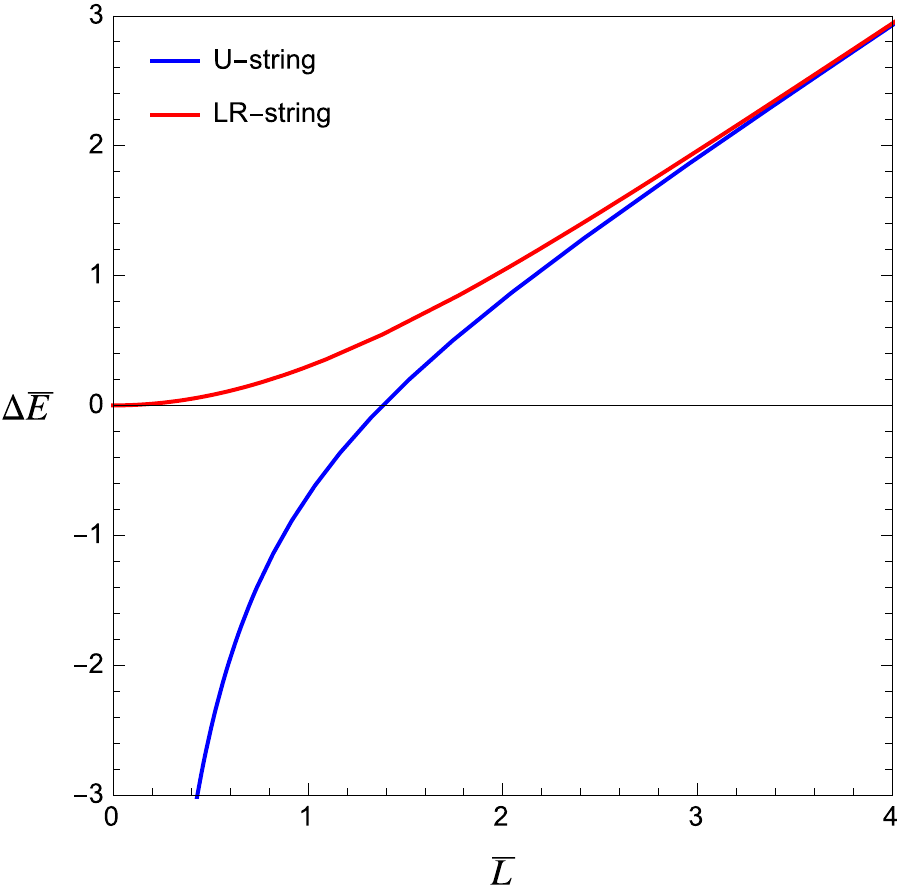}}
\vspace{-0.35cm}
\caption{The energy $\D\bar{E}$ of a quark-antiquark pair as a function of its separation $\bar{L}$. Configurations within the same boundary, dual to U-shaped strings, are plotted in blue, and Left-Right configurations, dual to strings which traverse the throat, are plotted in red.}
  \label{EvL_DOT}
\end{figure}

Following the same procedure described for strings in the thin-shell wormhole, we can calculate $\D\bar{E}$ in these backgrounds as a function of $\bar{L}$. 
This is shown in figure \ref{EvL_DOT}. For convenience, we have chosen $\r_0=0$ in (\ref{DOT_metric}). 

Qualitatively similar behavior for string probes as those studied in this letter has been observed in other AdS wormhole geometries. For example, in \cite{Taka}, holographic entanglement entropy \cite{RT1} was studied in a spacetime described by
\be
ds^2 = \ell^2 \lbr (1 + \r^2) \lp - d{t}^2 +d\vec{{x}}\cdot d\vec{{x}} \rp + \frac{d\r^2}{1 + \r^2} \rbr \, .\label{Ultima}
\ee
It is not difficult to see that, in the particular case when the entangling region on the boundary is the infinite stripe (equivalent to the rectangular Wilson loop that we have considered), the behavior of the holographic entanglement entropy is similar to that depicted in figure \ref{EvL_DOT}: For small $\bar{L}$, the behavior is the same as that for the pure AdS space, and for large $\bar{L}$ the result shows that it increases linearly. Our thin-shell wormhole solutions exhibit a similar tendency, with the difference being that cusp appear in the phase space due to the refraction effects produced when the strings go through the thin-shells. It is worth mentioning that, as in \cite{Taka}, the wormhole geometry (\ref{Ultima}) was also considered as an interesting scenario to study holographic entanglement entropy \cite{RT1, RT2, RT3}. Similar calculations can be performed in the thin-shell higher-curvature wormholes we constructed in this paper \cite{RT4, RT5, RT6}. This is matter of further investigation.

\section{Acknowledgments}

We are grateful to Alberto G\"uijosa for useful discussions. The work of M.C. and E.G. are partially supported by Mexico's National Council of Science and Technology (CONACyT) grant A1-S22886 and DGAPA-UNAM grant IN107520. The work of G.G. was partially supported by CONICET through the grant PIP 1109-2017.

%%%%%%%%%%%%%%%%%%%%%%%%%%%% BIBLIOGRAPHY %%%%%%%%%%%%%%%%%%%%%%%%%%%%%%%%%%%%%%

\end{document}